# Survey on Methods for Detection, Classification and Location of Faults in Power Systems Using Artificial Intelligence

**Juan A. Martinez-Velasco[1], Alexandre Serrano-Fontova[2], Ricard Bosch-Tous[3], Pau Casals-Torrens[4]**

1. Retired (formerly with Universitat Politècnica de Catalunya); juan-antonio.martinez@upc.edu
2. Reactive Technologies Ltd; advancedconsulting1990@gmail.com
3. Universitat Politècnica de Catalunya; ricard.bosch@upc.edu
4. Retired (formerly with Universitat Politècnica de Catalunya); p.casals@upc.edu

**Abstract:** Components of electrical power systems are susceptible to failures caused by lightning strikes, aging or human errors. These faults can cause equipment damage, affect system reliability, and results in expensive repair costs. As electric power systems are becoming more complex, traditional protection methods face limitations and shortcomings. The evolution of the traditional power system to the smart grid (SG) has changed the way to operate and protect power systems. The development of SGs calls for advanced fault diagnosis techniques to prevent undesired interruptions and expenses. Faults in power systems can occur at anytime and anywhere, can be caused by a natural disaster or an accident, and their occurrence can be hardly predicted or avoided; therefore, it is crucial to accurately estimate the fault location and quickly restore service. The development of methods capable of accurately detecting, locating and removing faults is essential (i.e. fast isolation of faults is necessary to maintain the system stability at transmission levels; accurate and fast detection and location of faults are essential for increasing reliability and customer satisfaction at distribution levels). This has motivated the development of new and more efficient methods. Methods developed to detect and locate faults in power systems can be divided into two categories, conventional and artificial intelligence-based techniques. Although the utilization of artificial intelligence (AI) techniques offer tremendous potential, they are challenging and time consuming (i.e. many AI techniques require training data for processing). This paper presents a survey of the application of AI techniques to fault diagnosis (detection, classification and location of faults) of lines and cables of power systems at both transmission and distribution levels. The paper provides a short introduction to AI concepts, a brief summary of the application of AI techniques to power system analysis and design, and a discussion on AI-based fault diagnosis methods.

**Keywords:** Artificial intelligence; Deep learning; Distribution system; Fault classification; Fault detection; Fault location; Machine learning; Metaheuristics; Power system; Transmission system

**Note:** For the sake of clarity, a list of acronyms (already defined here) has been attached to some tables of the paper.

## I. Introduction

Electrical power systems are very complex systems prone to different kinds of faults and failures. A fault can be defined as any abnormal system condition involving the failure of a power system component (e.g. overhead lines, insulated cables, transformers, generators, busbars). Faults in power systems are usually classified into two types: series (open conductor) faults, and shunt (short circuit) faults [1-3]. Series faults are characterized by unbalanced series impedances and can be identified by measuring phase voltages: any unusual voltage value increment indicates that an open conductor fault might have occurred. Shunt faults can be identified by observing phase currents: any unusual current value increment indicates that a short circuit fault might have occurred.

The causes of faults and failures in power systems are many. A power system component failure can be caused by human errors, adverse atmospheric conditions (e.g. lightning flashes) or adverse system conditions (e.g. overcurrents and overvoltages). A classification proposed in [4] considers the following categories of fault causes: natural phenomena, accidents, malicious attacks, and cascading faults. Assuming fault-free power systems is not reasonable, so predicting, locating and isolating faults and failed equipment is crucial for achieving high reliability and service continuity, avoiding system instability, or reducing interruption costs.

The transition to the smart grid (SG) is adding new metering, control, data acquisition and communication equipment. SGs integrate distributed energy resources (DERs), advanced sensing technologies, better controllers and communication technologies into the electrical grid to offer an intelligent manner to operate it with bidirectional power flow and self-healing capability [5-7]. Due to the complexity of the SG infrastructure, a multitude of faults and failures can emerge from different components and their interrelations, leading often to complex failure scenarios with potential cascading and disruptive effects. The increasing penetration of distributed generation (DG), microgrids (MGs), and load equipment in the existing power systems adds new challenges to conventional protection schemes, since most of the newer generation sources and loads use power electronics interfaces and allow bidirectional power flows, which lead to new protection issues (e.g. islanding problems, loss of coordination). The integration of more renewable energy resources (RESs) and storage systems, the addition of more monitoring and sensing equipment, as well as the implementation of fast communication networks, can improve system reliability and increase power quality, maintaining currents and voltages within safe ranges. However, such a number of different technologies introduces a growing number of failure points that, if not properly handled, can cause system instability and lead to cascading failures and blackouts. Therefore, an adequate strategy is required to detect, classify, localize, diagnose and isolate faults, and restore the system to normal operation as soon as possible. Some key features of SGs are their self-healing ability to locate and isolate disturbances, reduce fault frequency, reschedule resources to avoid critical situations, maintain the service continuity of the electric grid under any conditions, and shorten the time needed for outage restoration. It is then important to be able to classify and determine the faults and failures that can impact SGs, to look at the causes for preventive measures, and at consequences and countermeasures to counteract failure effects. The evolving concept of SG, along with the developments in information and communication technologies have contributed to the development of increasingly automated fault detection, isolation and restoration techniques [5-7]. Accurate detection, classification and location of faults and failures are crucial tasks to ensure enhanced performance by reducing system instability and long power outages.

Although protection schemes are different at transmission levels from those implemented at distribution levels, three stages can be distinguished in fault diagnosis of power systems irrespectively of the voltage level: fault detection, fault classification, and fault location; see, for instance, [8, 9]. The fault detection stage determines when a fault starts; it essentially separates events with faults from those without. The fault classification stage is turned on by the fault detection stage and aims at categorizing the fault type. After classifying the fault type, the fault location stage estimates the power system section and (if possible) the location of the fault. Figure 1 depicts a diagram of fault diagnosis in power systems. Major considerations for developing a fault diagnosis strategy are improving the fault detection procedure, locating the cause of the power outage, or deciding whether an online or offline location approach will be employed.

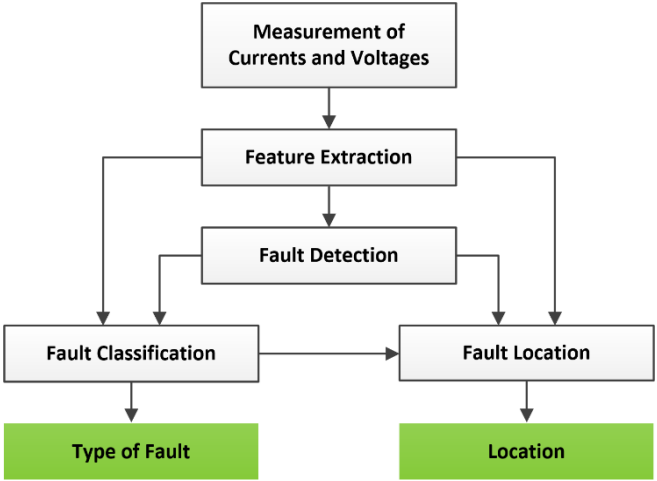

**Figure 1.** Fault diagnosis in power systems.

SGs need more accurate fault diagnosis algorithms, as more inverter-based generation sources, sensors, controllers and communication systems are added. The opportunity to implement more powerful and accurate methods for any stage of a fault diagnosis strategy increases as the grid observability and the performance of computer tools increases. As already mentioned, a fast fault diagnosis (i.e. fault detection, classification, and location) strategy can shorten power supply interruptions and ensure reliable operation of the power system. Although operators can use reports and complaints from the consumers, the increasing integration of monitoring and measuring equipment together with better communication networks allows quick fault detection. Two limitations for an efficient online fault detection have been the difficulty in obtaining data and the lack of computational capability and communication. The installation of intelligent electronic devices at different nodes of the grid and the use of synchronized global positioning system (GPS) sampling with high-speed broadband communications can circumvent these limitations.

SGs include, among others, distributed generation (e.g. photovoltaic, wind), energy storage, overhead lines and insulated cables, transformers, several types of loads, sensors, protective devices, and communication networks. Any part of component of the grid is prone to fail. For most of these components (e.g. transformers) a dedicated protection system is installed, and locating a fault is not an issue if a powerful supervisory and management system has been installed and is capable of fast reporting the failed equipment. An important concern is then the localization of faults on overhead lines and insulated cables. The goal of this paper is concentrated in fault diagnosis of faults and failures when the fault is located on a line or a cable at both transmission and distribution levels.

Many research documents highlighting the work carried out in fault detection, classification, and localization have been published [8-45]. The reviews cover either conventional approaches, AI-based techniques, or both.

Since the topology and the operation of transmission and distribution systems is different, the protection equipment and strategies implemented for each voltage level is also different. Even with the increasing integration of distributed resources at distribution levels, protection schemes are still different. In addition, the impact of disconnecting a transmission line/cable due to a fault/failure will be usually much higher than that of a distribution line/cable. These differences have led to different techniques of fault diagnosis:

- *Transmission*: The transmission system is based on a looped structure. Transmission lines are usually protected by relays installed at one or two ends that continuously monitor voltages and currents. The conventional methods often used in fault diagnosis of transmission lines are the traveling wave method and the impedance measurement-based method. The most frequent protection method is distance protection. The methods in this category can be further classified into two sub-categories: methods that use measurements from one terminal of the transmission line and methods that use measurements taken from both terminals [1, 2, 35, 46].
- *Distribution*: It is the most affected part of the power system since faults at these levels account for the majority of service interruptions to end-users. Traditionally, the default protection against faults in distribution networks is the overcurrent scheme. In contrast to transmission lines, the distribution networks are usually non-homogeneous, with branches and loads along a feeder, which make the fault location more difficult [1]. A very basic method of fault location uses visual inspection, which cannot be used if the fault is on an insulated cable. Conventional methods, proposed in literature or implemented in practice, use measured voltages and currents and may be divided into three categories [1]: methods based on traveling waves, methods using high frequency components of voltages and currents, and methods using fundamental frequency voltages and currents. The last method, also classified as impedance-based method, consists of calculating line impedances as seen from the line terminals and estimating distances of the faults. Impedance-based methods are popular among utilities, because of their ease of implementation.

The occurrence of high impedance faults (HIFs) is an interesting issue in fault diagnosis. An HIF is a type of fault that occurs on a network when an energized overhead conductor touches a high impedance surface or a poorly conducting surface (e.g. tree branches, rocks, sand, gravel, asphalt, concrete) [47-50]. A HIF typically produces a small fault current that is fluctuating by nature. The fault currents associated with HIFs are so low that they cannot trip the protection devices (i.e. they are insufficient to be detected by conventional overcurrent protection devices, such as fuses, reclosers, or protective relays). The existence of HIFs is one of the reasons that increases the percentage of high fault location errors of impedance-based methods. Traveling wave-based methods can be more accurate in finding HIFs; however, the required high sampling frequency, high implementation costs, and computational complexity make their practical application more difficult.

Many fault diagnosis techniques use signal processing methods such as fast Fourier transform (FFT), wavelet transform (WT), multiresolution analysis (MRA), and discrete wavelet transform (DWT) [42]. Some of these signal processing methods are actually part of many AI methods since they are extremely useful for feature extraction (see Figure 1) and to reduce the computational burden associated with AI-based approaches [8, 18, 35, 40].

AI techniques can facilitate performing accurate and efficient fault diagnosis. Actually, the application of artificial neural networks (ANNs) to classify and locate faults at both transmission and distribution levels is not new [51]. During the last two decades, many fault diagnosis methods using ANNs, support vector machines (SVMs), fuzzy logic, or genetic algorithms (GAs), just to mention a few techniques, have been proposed; see, for instance, [14, 19, 25, 33-37, 40-43]. Expert systems (ES) are another subfield of AI that were very popular in power system applications in the 1980s and 1990s [52-61].

This paper aims to present a survey of the different AI-based techniques used for fault detection, classification, and location estimation in power systems at both transmission and distribution levels, considering AC and DC systems, with or without DERs.

Artificial intelligence (AI) and machine learning (ML) are distinct concepts that fall under the same umbrella although they are often used interchangeably. AI is a branch of computer science that aims to mimic the ways that humans think in order to perform complex tasks, such as analyzing, reasoning, and learning. ML is a subset of AI that uses algorithms trained on data to produce models that can perform such complex tasks. To avoid misunderstandings, a definition of concepts related to AI is provided in Section II. The same section includes a subsection that also fixes the concepts related to *fault diagnosis* of power systems as applied in this paper. Once all concepts have been clearly defined, the scope of the paper is defined again.

Section III provides a short summary of applications of AI in power systems. Section IV is the core of the paper; it is divided into several subsections dedicated to reviewing and classifying fault diagnosis methods, and summarizing the state-of-the-art of fault diagnosis in transmission and distribution networks using AI techniques, including DC systems. The last two sections of the paper provide, respectively, a discussion on issues related to fault diagnosis and a short summary of the main conclusions of this work.

The limitations of the paper are discussed or mentioned in the subsequent sections, and it is important to emphasize that even the large list of references included in this document does not cover all aspects related to the main topics of the paper: a selection of papers has been made to limit the paper size, and fault diagnosis was limited to lines and cables; that is, fault diagnosis of transformers, wind turbines, or photovoltaic power plants, just to mention a few important power system components, has not been addressed. Finally, it is worth mentioning that the fault diagnosis of other system types (e.g., railway or marine power systems) is out of the scope of the paper.

## II. Concepts and Definitions on Artificial Intelligence and Fault Diagnosis

*2.1. Artificial Intelligence Concepts*

Alan Turing proposed in 1950 the idea of solving problems using machine intelligence [62]. The term *Artificial Intelligence* was coined by J. McCarthy, M.L. Minsky, N. Rochester, and C.E. Shannon in 1956 [63]. Although there is not a general accepted definition of AI, many definitions have been proposed. For instance, Russell and Norving define AI as "any entity that perceives its environment from sensors and acts in that environment from actuators can be described as an agent" [64]. An AI system may be seen as a system that simulates the functioning of the human mind, making computers, robots, and software think intelligently like humans. Since the birth of AI in the 1950s, the development of intelligent machines has been driven by a diverse array of approaches such as *statistical learning*, *knowledge-based systems* or *soft computing* [65]. Currently, there is a huge number of publications available for those interested in this subject. References [64-72] are a small sample of items that can be of help.

Actually, AI is an umbrella term that covers a variety of interrelated, but distinct, fields. Three main AI fields have been applied to power system studies [66]:

- **Machine Learning** (ML) is a subfield of AI focused on training algorithms with data sets to produce models capable of performing complex tasks [73-77]. Actually, an ML model is a computer program that can be used to recognize patterns in data or make predictions.

  Three types of ML algorithms can be distinguished:

  - ➢ **Supervised learning:** An algorithm is trained using data tagged with a label so that an algorithm can successfully learn from it. Training labels help the model know how to classify data in the desired manner. Two types of models can be distinguished in this group: classification and prediction. Classification models are used to assign test data into specific categories. Regression models use an algorithm to understand the relationship between dependent and independent variables, and can be used to predict numerical values from different data points.
  - ➢ **Unsupervised learning:** Unlabeled data are used to train an algorithm; the algorithm finds patterns in the data itself and creates its own data clusters. Unsupervised learning can be used to find patterns in data that are currently unknown.
  - ➢ **Reinforcement learning:** It is a technique in which positive and negative values are assigned to desired and undesired actions. The goal is to encourage programs to avoid the negative training examples and seek out the positive learning how to maximize rewards through trial and error.

  Some authors add other types:

  - ➢ **Semi-supervised learning:** It uses a mix of labeled and unlabeled data to train an algorithm. In this process, the algorithm is first trained with a small amount of labeled data before being trained with a much larger amount of unlabeled data.
  - ➢ **Ensemble methods:** They make use of several ML algorithms to improve the performance that can be achieved with the use of a single algorithm. Ensemble learning constructs a set of hypotheses generated by several base learners that are used together to solve a single problem and provide better generalizability than individual base learners.

  Before training an ML algorithm, the so-called hyperparameters must be set; they act as external guides that inform the decision process and direct how the algorithm will learn. The number of branches on a regression tree, the learning rate or the number of clusters in a clustering algorithm are examples of hyperparameters. Hyperparameters train and direct the algorithm; parameters begin to form in response to the training data. These parameters include the weights and biases formed by the algorithm as it is being trained.

  Figure 2 depicts a list of ML techniques classified according to the categories mentioned above; see also [24]. For a definition and description of these techniques see the recommended literature [64, 70-76].

- **Metaheuristic methods** are a type of algorithm characterized by their ability to solve optimization problems by mimicking natural phenomena or human intelligence [77, 78]. They are used to find approximate optimal solutions to complex and highly nonlinear optimization problems for which no deterministic approach is able to handle in an acceptable amount of time. The *particle swarm optimization* (PSO), the *ant colony optimization* (ACO), the *genetic algorithm* (GA), the *tabu search* method or the *simulated annealing* method are some of the most popular algorithms. A significant effort has also been carried out to review this group of methods, see references [79-86]. It is worth mentioning that reference [84] provides a list of more than 500 metaheuristics algorithms, and it is not complete since some additional methods have been proposed; see, for instance, [87-91].

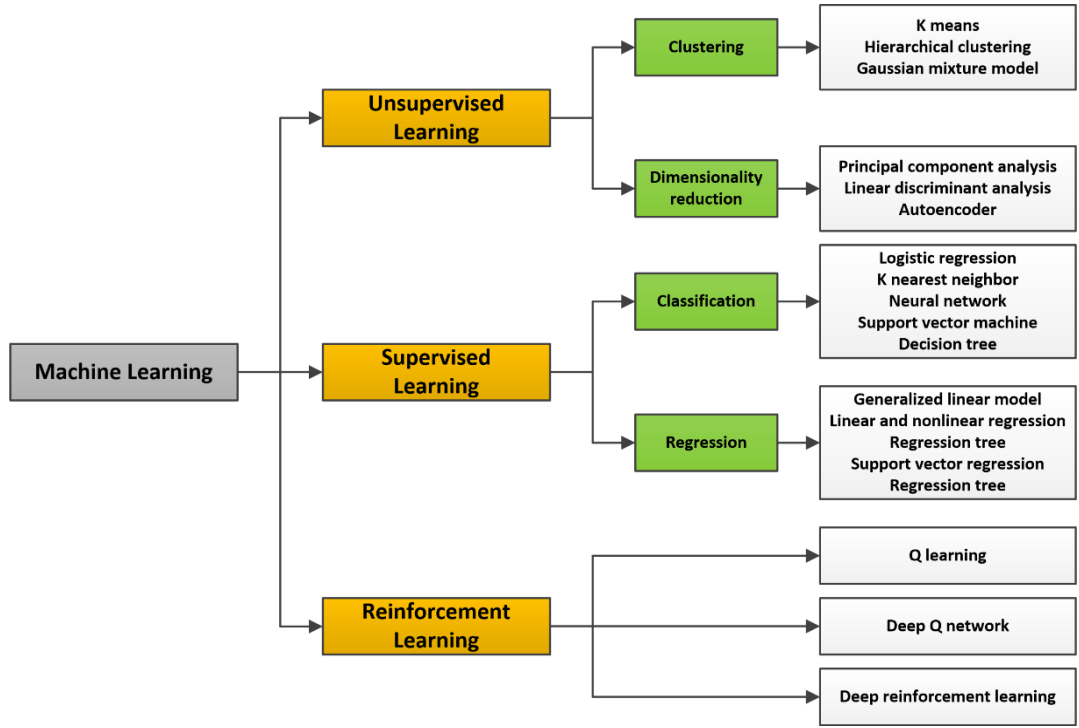

**Figure 2.** Machine learning techniques (based on [24]).

- **Rule-based systems**, also referred to as **expert systems** (ESs), are a group of techniques that allow the direct integration of human knowledge. An expert system is a computer software that emulates the process used by human experts when they solve problems [92]. By developing a set of if-then rules, the system is able to decide based on the rules given by an expert. Besides the Boolean logic, fuzzy logic has also been used in rule-based systems; its main advantage is the description of variables and relations in human linguistics. A fuzzy system normally consists of three basic parts [93]: (1) *fuzzification*, where the input signals are mapped onto a fuzzy membership function using a membership degree; (2) *inference*, where the calculated degrees of membership are integrated into IF-THEN fuzzy rules; (3) *defuzzification*, which creates an output signal that the physical system is able to handle. For more information on this AI subfield, see [94-98].

Figure 3 shows a diagram with the different AI groups and subgroups mentioned above.

**Deep Learning** (DL) is a subset of ML, in which artificial neural networks (ANNs) are used to perform more complex reasoning tasks without human intervention. The main differences between DL and ML are the neural network architecture, how each algorithm learns, and how much data each type of algorithm uses. Non-deep traditional ML models use simple neural networks with one or two layers, DL models use three or more layers (typically hundreds or thousands of layers) to train the models. Figure 4 schematizes the relationship between AI, ML and DL. While supervised ML models require structured labeled input data, DL models can use unsupervised learning. With unsupervised learning, DL models can extract the characteristics, features and relationships they need to make accurate outputs from raw unstructured data. DL automates much of the feature extraction piece of the process, eliminating some of the manual human intervention required. In general, a DL model requires more data values than an ML model to improve accuracy. A significant number of publications is available to readers interested in this subject; see, for instance, [99-107].

**Natural Language Processing** and **Robotics** are two fields also covered by the AI umbrella.

*2.2. Fault Diagnosis Concepts*

A first aspect to be clarified is the distinction between failure and fault, two terms that are often indistinctively used.

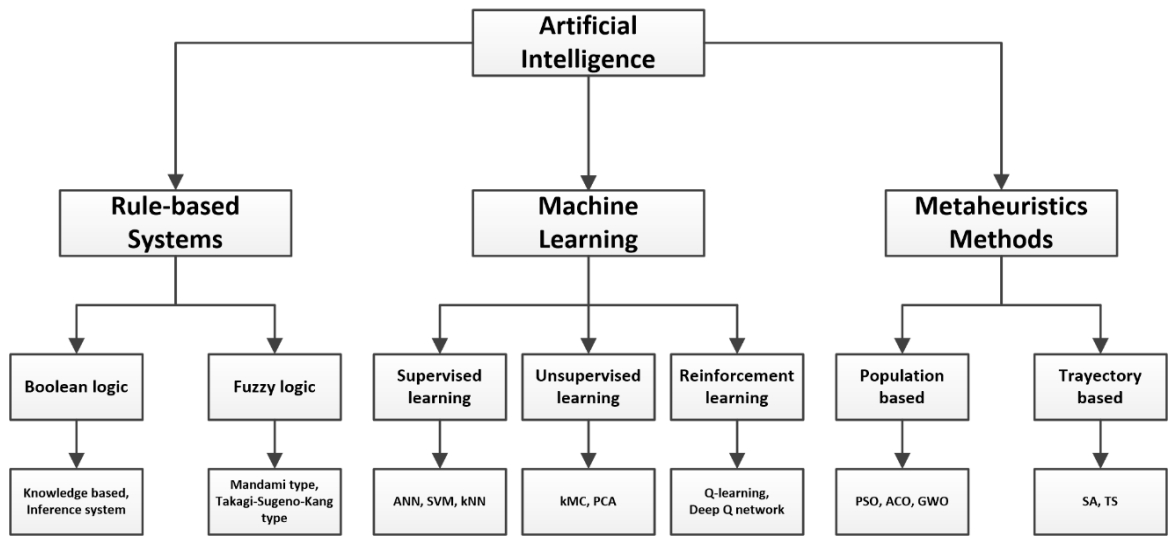

**Figure 3.** Categories of AI.

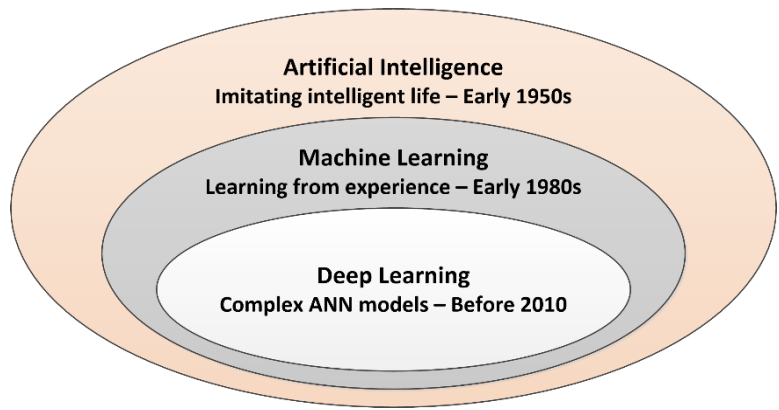

**Figure 4.** Artificial Intelligence vs Machine Learning vs Deep Learning.

A **failure** of a power component may be defined as an inability condition of that component to perform required functions on demand due to one or more defects. A failure of a component is, therefore, the termination of its ability to perform a required function [108]. That is, a component has suffered failure when it is no longer capable of fulfilling one or more of its intended functions; this does not mean that it is completely unable to function.

The cause that initiates the failure is the *failure cause*, while the physical process that leads to the failure is the *failure mechanism*. A failure can cause loss or damage to other power components. The final consequence of component failures is known as *outage* when a disruption occurs in either generation, transmission or distribution of electricity, causing a loss of power to some or all customers in the system [108].

A component failure might result in a fault. A **fault** in a power system can be defined as an abnormal electrical condition which interferes with the normal power flow and may affect the capability of the power system to fulfill its function [108]. A fault is a defect in the electrical circuit due to which current is diverted from the intended path and that can result in damage to other power components. A fault is a defect that causes a component to fail or act abnormally and may lead to a failure.

Note that the two definitions are circular: a failure may lead to a fault, a fault may lead to a failure. Actually, this is not contradictory. Consider the following example: an overvoltage causes an insulation breakdown to a non-properly protected underground cable (i.e. the overvoltage is the cause, the insulation breakdown the mechanism of the failure); this failure causes a fault that increases significantly the system current, which in turn may cause a failure to other components affected by the overcurrent.

When a fault occurs in a power system, a protection system scheme is needed to isolate the faulty part from the healthy part. If the protection system acts and a component is switched off, an outage in the power system occurs. A distinction can be made between forced outages (caused by failures of components switched off by the protection system, or disconnected manually by the system operator immediately) and planned outages (due to maintenance activities) [108]. Faults cause interruption to electric flows, equipment damages, and even death of humans and animals. The occurrence of a fault should be followed by the quick isolation of the faulty zone and the service restoration to those zones affected by the fault and to which the service can be reestablished. Only after reparation or replacement of the failed component, restoration can be then performed to reestablish the service to the entire system.

To clarify the scope of the paper, consider the following definitions [29]:

- fault detection is the task of recognizing the occurrence of a fault;
- fault classification is the identification of the fault type;
- fault isolation is the process of isolating the faulty part of the network after a successful detection;
- fault location is the task of localizing the fault (i.e. branch, zone, location point);
- service restoration is the process aimed at returning the system to normal operating conditions; that is, a task carried out to restore the service in healthy zones after detecting and isolating a permanent fault in the system.

In this paper, **fault diagnosis** is the combination of fault detection, classification and location. Neither isolation nor restoration are covered here. Methodologies and practices for service restoration depend on utilities and the voltage level of the affected system; see, for instance, [109, 110].

As mentioned in the Introduction, faults in power systems can be classified into two groups: series and shunt. Figure 5 shows a more complete classification of faults in power systems; see, for instance, [21, 37]. Faults can also be classified as temporary and permanent. Temporary faults are faults that occur for a certain period of time and can be cleared with or without interrupting operation (i.e. depending on their nature and duration, they can or cannot require the involvement of protective devices). Permanent faults will require the involvement of the protection system to interrupt the service within the faulty zone. A lightning stroke can cause a temporary fault in an overhead line without creating permanent damage; in addition, depending on the conditions within the impacted line, the fault can force or not the operation of the protection system. By default, a fault in an insulated cable is, irrespective of its cause, always permanent.

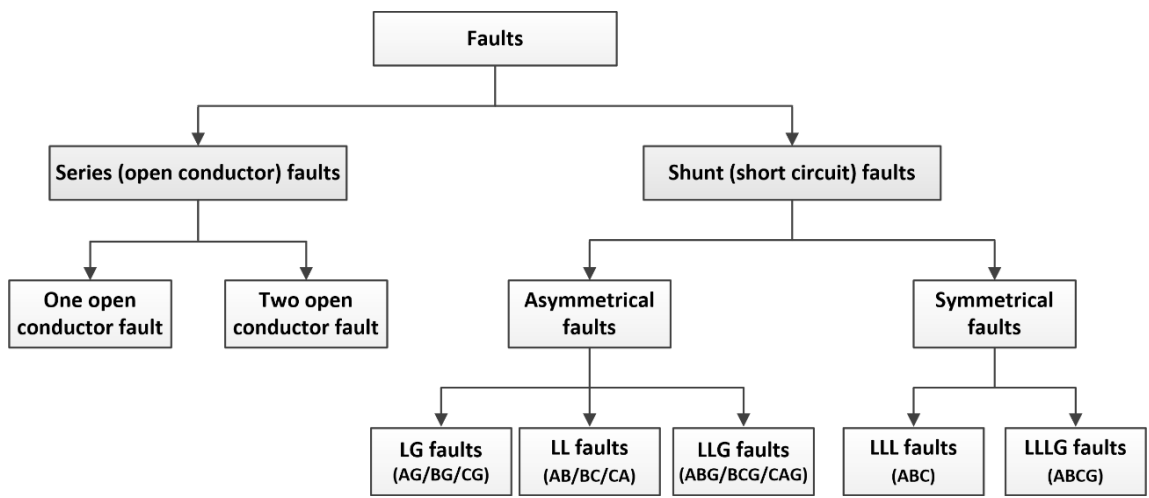

**Figure 5.** Classification of faults in power systems.

A fault in a distribution system is de-energized by opening and locking out breakers or reclosers. This lockout creates an outage for the faulted zone (i.e. customers downstream of the faulted zone, and also customers upstream of the faulted zone, in case of miscoordination). Distribution companies are implementing **FLISR** (fault location, isolation, and service restoration) solutions, also known as FDIR (fault detection, isolation, and restoration); see, for instance, [30, 111, 112]. FLISR automatically restores power to as many customers as possible, as quickly as possible. Without FLISR, an outage must be handled manually, often resulting in a longer outage larger. Deploying FLISR makes it possible to improve the coordination of switching devices and extend their utility into post-fault restoration; see Figure 6. FLISR technologies involve automated feeder switches and reclosers, line monitors, communication networks, supervisory control and data acquisition (SCADA) systems, remote terminal units, and data processing tools.

These technologies work in tandem to automate power restoration, reducing both the impact and length of power interruptions automatically as quickly as possible. In addition, the fault isolation feature can help crews locate the trouble location more quickly, resulting in shorter outage durations for the affected customers. A *fault passage indicator* (FPI) is a device used in distribution systems to detect and locate faults on power lines. When a fault occurs, the FPI detects the resulting change in current and transmits a signal to a central control system. When properly applied, FPIs can decrease operating costs and reduce service interruptions by quickly identifying the failed zone.

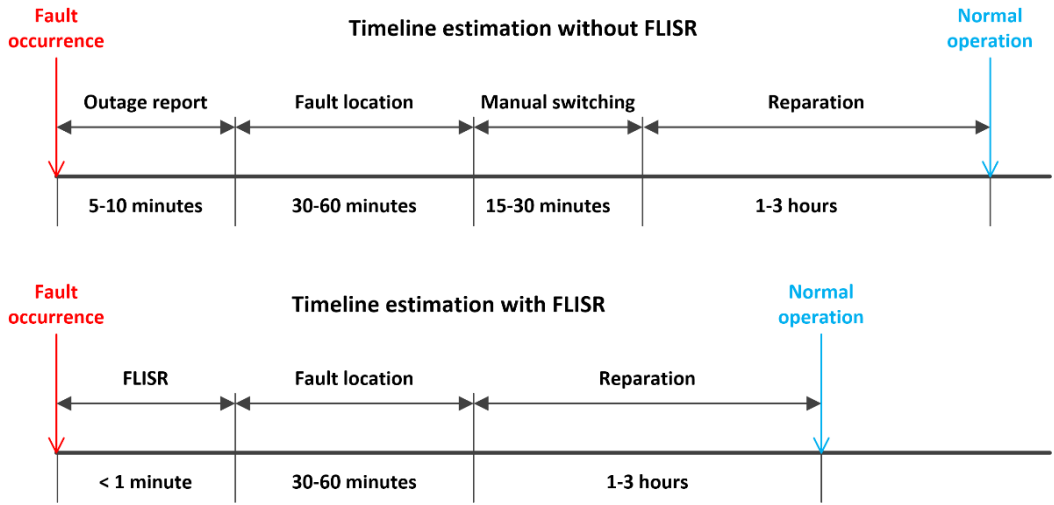

**Figure 6.** Service restoration without and with FLISR solution.

*2.3.     Scope of the Paper*

Once the concepts to be used and analyzed in this paper have been defined, it is worth defining again the scope of this paper.

As mentioned in the previous subsections, a service interruption in a power system can be caused by a failure in many system components. However, many of these components are crucial for the operation of the system and they are well protected. Consequently, failures in any of these components will usually lead to the operation of its corresponding protection system, and locating the fault position should not be a serious issue. Of course, power systems are very complex and the location of the faulty component will not always be obvious. However, this task can be carried out, for instance, by an expert system developed for analyzing alarms. In any case, such location task would not be based on complex mathematical analyses. A different scenario is presented when the failure is on an overhead line or an insulated cable; the faulted position cannot be then always located by a fast visual inspection, mainly if it is inside an insulated and underground cable.

According to the concepts defined in the previous subsection a complete fault diagnosis might include the identification of the faulted section. Given the design of protective systems, this could be an issue for distribution systems for which the faulted section is not protected at any of its ends, and therefore the section is not isolated in case of fault. In such cases, the fault location and section identification must be carried out simultaneously.

The goal of this paper is to provide a bibliographical survey of AI-based techniques proposed to date for detection, classification and location of faults in overhead lines and insulated cables, at both transmission and distribution levels, being lines and cables operating in either AC or DC regimes.

## III. Application of Artificial Intelligence Techniques to Power System Studies

The electric power grid is a complex and dynamic system that is responsible for delivering electricity to industries, businesses, and homes. The traditional power grid has been facing numerous challenges that include inefficiencies, lack of flexibility, and aging infrastructure. Traditional power systems have historically relied on a small number of centralized and large power generation sources (e.g. hydropower, nuclear, fossil-based power generation plants); the generated energy is transmitted using large transmission networks and delivered to consumers through distribution networks (Figure 7). This system is passive with unidirectional power and communication flows.

This system is currently undergoing a transition to a more flexible and efficient SG that uses a two-way communication to optimize power delivery, improve efficiency, and support the integration of renewable energy sources; see Figure 7. Smart technologies that are being implemented in modern power systems include advanced metering infrastructure, automated control systems, flexible power-electronics-based converters and low-latency communication networks [5-7]. They enable utilities to better forecast needs and make necessary adjustments in order to ensure reliable delivery of electricity: advanced sensors provide real-time information about energy production and consumption and allow for smarter grid management; smart meters provide customers with detailed information about their energy usage and help them to make informed decisions about how to use electricity; new automation capabilities facilitate to quickly respond to power generation intermittencies that could affect the power system and to automatically adjust electrical appliances in response to changes in the network. The integration of electric vehicles is also facilitated by the implementation of the SG.

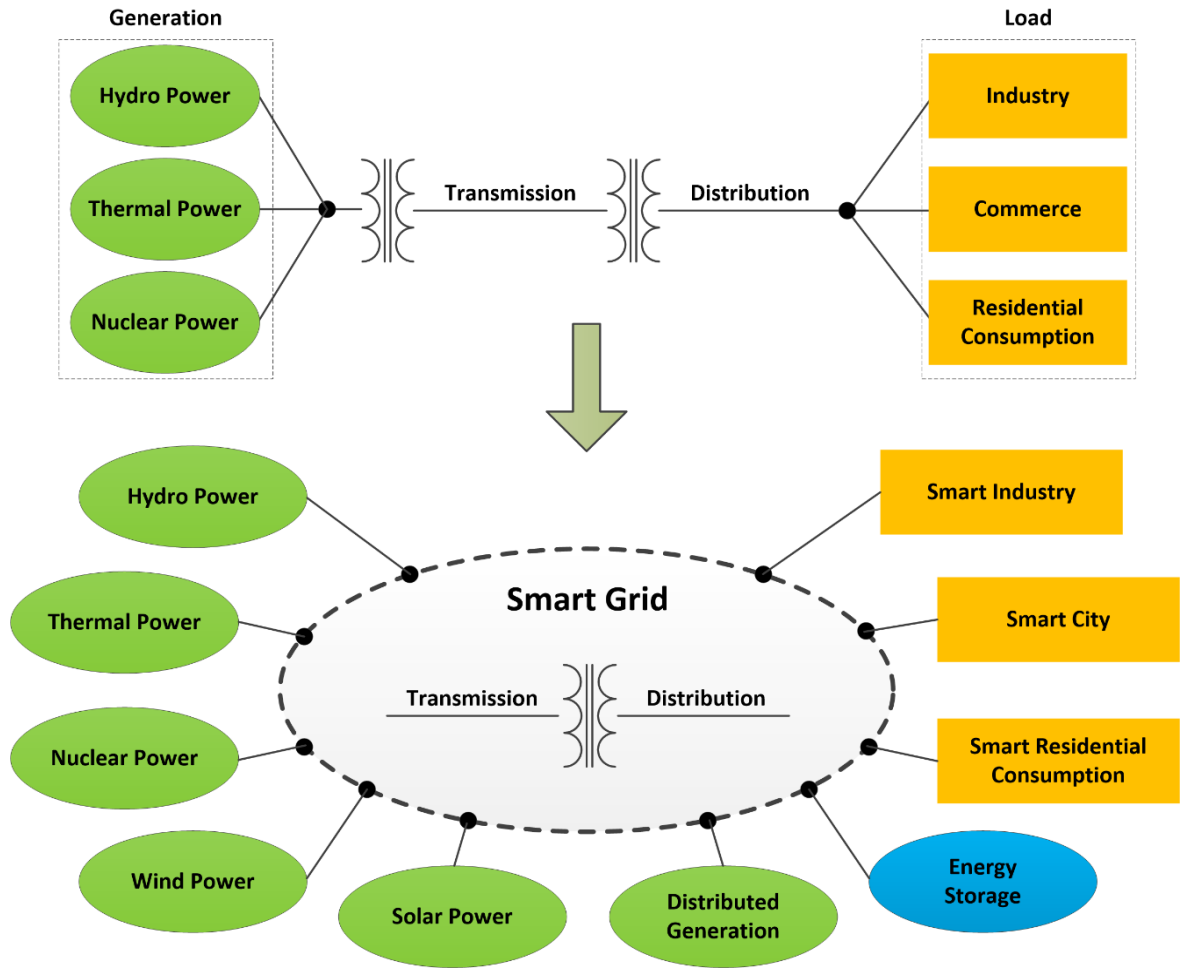

**Figure 7.** The transition from the traditional power grid to the smart grid.

Technologies used to monitor the delivered energy allows for two-way communication between utilities and consumers, providing them with insight into their energy use. The integration of renewable power generation at a local level improves environmental performance and allows for two-way flow of electricity by enabling electricity to be transmitted from both sources (utilities and consumers). Energy prosumers (those who consume the energy from the grid and also produce their own energy) can connect directly with the electric network and use the energy they generate efficiently. A SG equipped with sensors can quickly detect faults within the grid infrastructure, and resolve any outages in a timely manner. The introduction of these technologies offers utilities the ability to integrate storage and renewable generation systems into the existing distribution system and provide an increased level of flexibility by changing energy consumption patterns in response to demand. SGs also allow for better control of power flows, enabling better management of intermittent renewable sources. In addition, communication technologies can be used to enhance the efficiency of the grid by enabling shorter response times and higher levels of accuracy in monitoring system performance.

A SG is characterized by its active nature, a result from the two-way power and information flow enabled by the integration of distributed energy resources and the implementation of low-latency communication networks [5-7].

One important component of the SG is the microgrid (MG) [113], which is basically a cluster of loads and micro-generation sources (e.g., wind turbines, micro-turbines, photovoltaic plants, fuel cells), operating as a single controllable system that delivers power to local consumers. A MG can operate in two modes: connected to the main power system (*coupled mode*), and isolated from the power system (*island mode*). The interconnection of MGs with the utility grid is usually made through a power electronic converter, which plays a vital role in regulating the power flow between the MG and the main grid. MGs face severe challenges regarding power quality and security. In order to satisfy the power quality standards and to ensure the smooth operation of the power system during and after the grid connection, a robust control strategy is essentially required.

The SG exhibits an increased level of complexity with respect to the traditional power grid. Its study demands more complex and powerful algorithms that need to process vast amounts of data due to the increased observability provided by the multiple sensors installed in the grid or the uncertainty caused by intermittent renewable sources. As power grids become more complex, the application of advanced techniques in the analysis, design or management of power systems become a necessity. The steady improvement of hardware and software capabilities have empowered the development of various solutions based on AI.

The incorporation of DG poses several challenges to protection systems, particularly in distribution networks: false tripping; miscoordination of protection devices; variable short-circuit levels (synchronous generators can sustain high fault current for rather long periods of time, but fault currents generated by induction generators decay very quickly). The limited short-circuit currents that can be caused by converter-interfaced sources may prevent the normal operation of traditional overcurrent protection devices. In addition, unintentional or unplanned islanding can occur when the distribution system is disconnected from the main power grid; such scenario can cause damage to system equipment in the isolated section (and even be dangerous for people). Therefore, protection schemes that could detect generation islands are required. Protection challenges associated with MGs can also arise with the addition of DG sources. Therefore, the transition towards a smart grid with penetration of DERs and MGs leads to new protection schemes and devices. They also affect fault diagnosis tasks. For more details on the impact that DGs and MGs can have on power system protection and fault location methods, see [114-117].

The presence of FACTS devices poses another challenge to the system protection of transmission lines since their impedance is constantly changing: a FACTS device can inject either a variable series voltage or a variable shunt variable current along the line and introduce harmonics as well as non-linearities that can affect the line impedance value seen by protective relays. Therefore, protection and fault diagnosis methods become more complex than for simple uncompensated lines. For more details on the impact that FACTS devices have on protection see [27, 38, 118].

The number of research papers, books, technical reports and case studies related to AI application to power systems is quickly increasing. AI is already playing and will continue playing a crucial role in future power systems by introducing advanced techniques useful for power system studies. AI-based solutions are now being used in power flow optimization, fault detection, predictive maintenance, demand response, energy management, or forecasting of power demand and renewable energy, just to mention a few. To date a huge number of papers, reports and books analyzing and solving issues related to the power systems has been published. Since even a short survey of AI-based solutions would be out of the scope of this paper, a list of review papers has been selected to illustrate the scope of AI applications in this field; see references [119-211]. Some books on this subject are listed in references [212-226]. Note that initially the AI solutions applied to power systems were basically ESs and ANNs. It was after 2010 when ML and DL techniques became predominant. Many of these reviews include a section on fault detection and design; see, for instance, [145, 150, 151, 155, 159, 176, 178, 180, 194, 199, 204]. This is the main topic of this paper and will be deeply covered in the next section.

A selected number of review papers published since 2021 is presented in Table 1. Although several of these papers are connected to the same topic, in general the approaches used by the authors for analyzing a single topic (e.g. power system protection, power system operation) are different, so it is recommendable to consider all of them.

The concept Explainable Artificial Intelligence (XAI) has been developed to improve the explainability of AI models and better understand their output [227-232]. Traditional AI models are often seen as black boxes whose outputs are opaque and/or difficult to interpret. XAI refers to AI models that provide insights into how they make decisions, and allow users to understand them. Therefore, XAI brings benefits that can enhance both the development and deployment of AI models. For instance, when applied to fault detection, a XAI model can detect what sensor data deviate from normal values, and help maintenance crews to clarify the causes of an anomaly. Reference [182] provides a review of XAI applications to energy and power systems (e.g., transient stability assessment, fault diagnosis, load and photovoltaic power generation forecasting, grid control). XAI has, however, limitations: a trade-off between performance and interpretability (an accessible explanation might not completely represent the decision-making process and potentially mislead users), computational complexity (some XAI methods involve significant computational overhead), fidelity (an approximated explanation might not accurately reflect the actual model behavior), bias (a XAI model can reproduce the bias present in the training data), lack of standardization (there is no standard metric for what is a good explanation).

Many of the papers listed in the table cope with a single field (e.g., protection, energy management, stability, operation and control). Although the goal of some references was to cope with all power system studies, actually none covered all of them. Figure 8 shows a framework with some of the most common applications of AI techniques in power systems. It is worth mentioning that the figure shows just one way of classifying the many studies in power systems to which AI techniques have been applied to date and it is based on a classification proposed in [226] (see Chapter 3). As an example, Table 2 shows a summary of the applications of AI techniques in energy management systems. The table shows a very short description of some energy management services (EMSs) fields and some of the AI techniques used in their study and solution.

**Table 1.** – Selected reviews on application of AI techniques to power system studies.

| Ref. | Authors | Year | Topics |
|---|---|---|---|
| [211] | Mishra and Singh | 2025 | Overview of AI applications to power system protection |
| [210] | Judge et al. | 2024 | Overview of AI applications to SG integration and optimization |
| [209] | Alhamrouni et al. | 2024 | AI applications to power system stability, control, and protection |
| [207] | Zahraoui et al. | 2024 | Overview of AI applications to resilience in power systems and MGs |
| [206] | Hallmann et al. | 2024 | Overview of AI applications to power system operation |
| [205] | Porawagamage et al. | 2024 | Overview of power system protection and control |
| [204] | Chen et al. | 2024 | Overview of AI applications to power system studies |
| [203] | Akter et al. | 2024 | Microgrid optimization using meta-heuristic techniques |
| [202] | Lee et al. | 2024 | Overview of ML applications to energy management systems |
| [201] | Heymann et al. | 2024 | Overview of AI applications to power system studies |
| [200] | Pandey et al. | 2023 | AI applications to power system operation, control and planning |
| [199] | Akhtar et al. | 2023 | Overview of DL applications to power systems |
| [197] | Ruan et al. | 2023 | Overview of DL applications to cybersecurity studies |
| [196] | Mehta et al. | 2023 | Overview of AI applications to green energy studies |
| [194] | Strielkowski et al. | 2023 | Overview of ML applications to predictive analysis of power systems |
| [192] | Markovic et al. | 2023 | Overview of ML applications to distribution system studies |
| [191] | Chung and Zhang | 2023 | Overview of AI applications to distribution system studies |
| [190] | Wang et al. | 2023 | Overview of ML applications to power system resilience studies |
| [189] | Franki et al. | 2023 | Overview of AI applications to power system studies |
| [187] | Entezari et al. | 2023 | Survey of AI applications to energy systems studies |
| [185] | Aminifar et al. | 2022 | Overview of AI applications to power system protection |
| [184] | Ahmad et al. | 2022 | Overview of AI applications to energy systems |
| [181] | Chen et al. | 2022 | Survey of RL applications to power system studies |
| [180] | Jafari et al. | 2022 | Survey of DL applications to distribution automation studies |
| [178] | Forootan et al. | 2022 | Overview of AI applications to energy systems |
| [174] | Kumbhar et al. | 2021 | Overview of ML applications to power systems |
| [173] | Stock et al. | 2021 | Overview of AI applications to distribution system operation |
| [172] | Barja-Martinez et al. | 2021 | Overview of AI applications to distribution systems |
| [171] | Donti and Kolter | 2021 | Overview of ML applications to energy systems |
| [170] | Miraftabzadeh et al. | 2021 | Survey of ML applications to power system studies |
| [169] | Aslam et al. | 2021 | Survey of DL applications to MG studies |
| [168] | Massaoudi et al. | 2021 | Survey of DL applications to smart grid studies |
| [167] | Omitaomu and Niu | 2021 | Survey of AI applications to smart grid studies |
| [165] | Khodayar et al. | 2021 | Survey of DL applications to power system studies |
| [164] | Perera and Kamalaruban | 2021 | Survey of RL applications to energy system studies |
| [163] | Wu and Wang | 2021 | Overview of AI applications to MG operation and control |
| [158] | Feng et al. | 2021 | Overview of ML applications to power system studies |

AI: Artificial Intelligence, DL: Deep Learning, MG: microgrid, ML: Machine Learning, RL: Reinforcement Learning, SG: Smart grid.

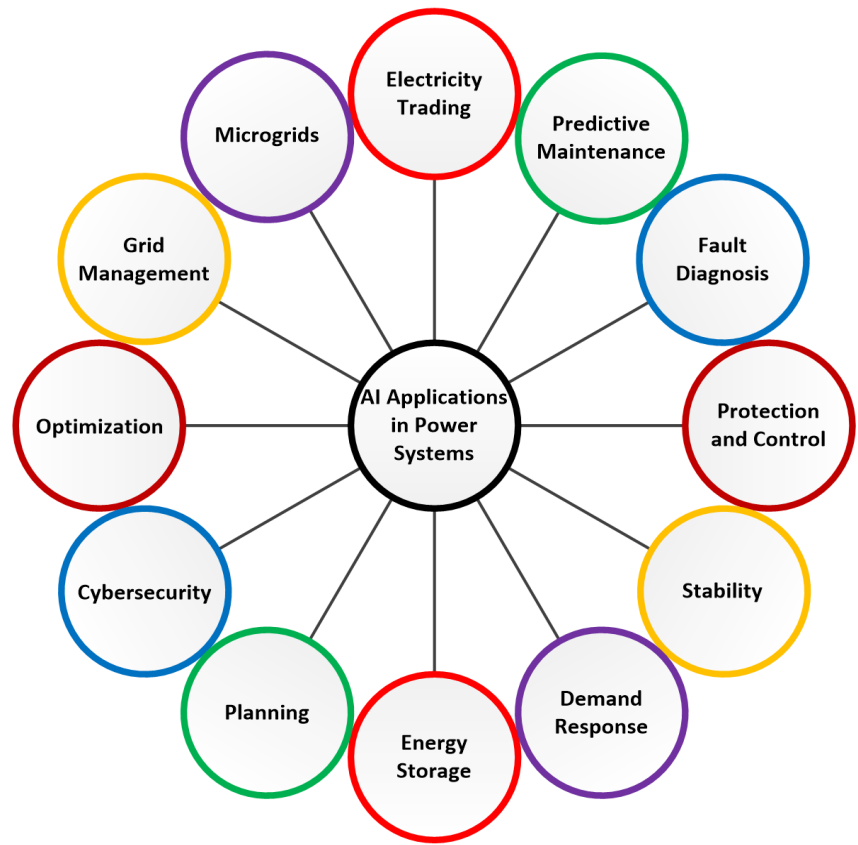

**Figure 8.** Some important AI applications in power systems.

**Table 2.** – Example of AI applications in EMSs (based on [209] and [210]).

| Application | Description | AI Techniques |
|---|---|---|
| Demand Forecasting | Predicts energy demand to optimize power generation and distribution. | ML (FLR, SVM, RF, KMC) DL (RNNs, LSTM, DNNs) |
| Grid Management | Real-time monitoring and control of power grids to detect and predict faults, manage loads, and optimize electricity flow. | ML (FLR, SVM, KMC, QL) |
| Renewable Energy Integration | Predicts the generation capacity of renewable sources based on weather conditions, improving grid stability and optimizing renewable use. | ML (FLR, SVM, RF, KMC) DL (RNNs, LSTM, DNNs) |
| Energy Storage Management | Optimizes charging and discharging cycles of energy storage systems, prolonging battery life and reducing costs. | ML (SVM, KMC, QL) DL (RNNs, LSTM, DNNs) |
| Predictive Maintenance | Analyzes sensor data to predict equipment failures before they occur, reducing downtime and maintenance costs. | ML (RF) DL (LSTM, DNNs) |
| Energy Efficiency | AI-driven systems in buildings and industries optimize energy use by learning consumption patterns and implementing efficiency measures. | ML (FLR, SVM, RF, KMC) |
| Fault Diagnosis | Quickly identifies and diagnoses faults in the power system, enabling faster responses and preventing equipment damage. | ML (SVM, RF, KMC) DL (RNNs, DNNs) |
| Energy Trading | Forecasts prices, optimizes trading strategies and manages risks in energy trading platforms. | ML (FLR, SVM, RF) DL (RNNs, LSTM, DNNs) |

ML: Machine Learning, DL: Deep Learning, FLR: Fuzzy Linear Regression, SVM: Support Vector Machine, RF: Random Forest, KMC: K-means Clustering, LSTM: Long Short Term Memory, DNN: Deep Neural Network, RNN: Recurrent Neural Network, QL: Q-Learning.

# IV. Fault Diagnosis of Power Systems Using Artificial Intelligence Techniques

*4.1. Introduction*

The goals of this paper were presented in previous sections after providing a short introduction to the main topics and a brief description of the main concepts. This subsection presents a more detailed description of methods aimed at detecting, classifying and locating faults in lines and cables working at any voltage level, considering both AC and DC operation.

Figure 1 depicted a diagram of fault diagnosis, as assumed in this work. According to the figure the tasks to be performed to classify and locate a fault are as follows: (1) measurements of voltages and currents, (2) extraction of features from the measured signals; (3) fault detection; (4) fault classification; (5) fault location.

The techniques in fault diagnosis can be classified into two distinct categories [11]: model-based and process history-based; see Figure 9. Model-based methods use a mathematical model to describe the process system; the description is derived from the underlying physics of the process and can be quantitative or qualitative. Process history-based methods do not use an explicit analytical relationship to define the physical system, they use instead a large set of experimental data set from which they derive a function relating input and output. The relationship between inputs and outputs can be statistical (e.g., regression) or non-statistical (e.g., ANN, SVM).

In the rest of this section, it is assumed that voltage and current measurements are available to perform the subsequent tasks; for a review of sensor technologies used in fault diagnosis for power systems see reference [233]. A short introduction to those tasks is presented below.

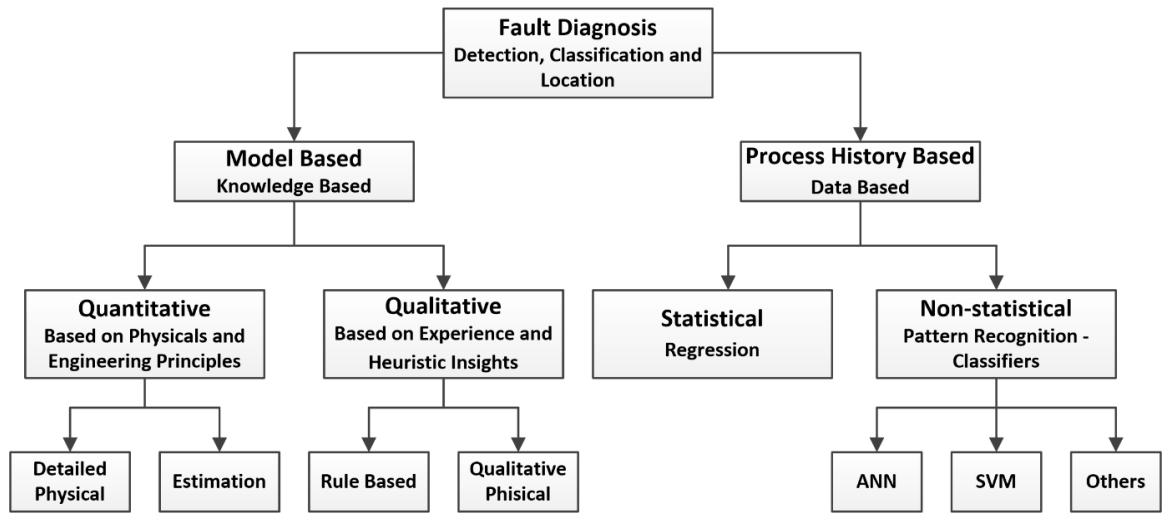

**Figure 9.** Categories of fault diagnosis (based on [11]).

**Feature extraction:** Although measured current and voltage signals contain all the required information for a fault diagnosis, it is not easy to apply a set of rules to sampled signals; a feature extraction is generally used to disclose the information required by the subsequent tasks. Detecting, classifying and locating a fault are much more easily performed by using an appropriate and reduced list of values derived from the sampled waveforms: dimensionally reduced data will boost the performance of any algorithm and provide faster results. The following classification of techniques aimed at extracting features from sampled signals was presented in reference [9].

1. **Transform methods**: The frequency characteristics of current and voltage signals during a fault change with time, and they can be very useful for detecting, classifying and locating a fault. A variety of methods used to analyze frequency characteristics of time-domain signals have been proposed. Some of the most popular transform methods are the Fourier transform (FT), the wavelet transform (WT), and the S transform (ST). Since time-domain signals and frequency-domain coefficients are both discrete, in practice the most popular approaches use this condition for feature extraction; they are the discrete Fourier transform (DFT), the fast Fourier transform (FFT), or the discrete wavelet transform (DWT).
2. **Modal transformations**: They transform/decouple three-phase quantities into components that can be used to further characterize fault types or to obtain detection and location indices. A list of popular transformations includes the Clark transformation, the Clarke-Concordia transformation, and the Karrenbauer transformation.
3. **Dimensionality reduction**: This approach maps the data from the original high-dimensional space onto a low-dimensional subspace in which the variance of the data can be best accounted. The reduction is usually performed by means of principal component analysis (PCA), and can be combined with other methods.

4. **Other methods**: To reduce the computational burden associated with the above listed methods, other approaches have been proposed. They are based on the RMS values of phase and zero sequence currents, the normalized ratios of maximum absolute values of currents for two different phases, or the ratios of phase angle differences between phases plus the ratio of zero sequence current amplitude to positive sequence current amplitude [9]. Mathematical morphology is another option that is being adopted as a feature extraction technique for detection and classification of faults [234]. For a comparison of feature extraction techniques, see this last reference.

**Fault detection:** This task is aimed at determining whether a fault exists or is developing. A thorough list of techniques for the detection of faults and failures in SGs is provided in reference [31].

The first obvious approach for detecting a fault in a line or a cable is based on a threshold value of a current parameter. Since high current values in power systems are associated with low voltage values, a more sophisticated approach can include a threshold value of a voltage parameter that would confirm the existence of a fault current. Although fault detection is generally performed using the information derived from the feature extraction, many modern algorithms are proficient enough to avoid this stage; for instance, when the classifier or the locator is capable of distinguishing between faulty and non-faulty states.

Detection methods independent of the classification methods use negative sequence components or parameters extracted by using a wavelet transform. Several methods have also been proposed for detecting high impedance faults using these approaches; see [9].

This task has been the subject of many review papers; see, for instance, references [23, 145, 157, 167, 204, 206, 211]. High impedance fault detection techniques on distribution networks have been reviewed in [49, 50, 235].

**Fault classification:** Figure 5 presented a classification of faults in power systems with two categories: series and shunt faults. Faults can also be static or evolving [236]. Evolving faults, also known as multi-stage faults, are those faults that begin as one type and change over time to another type; for instance, a single-line-to-ground fault that becomes a phase-to-phase with ground contact fault.

References [237] and [238] proposed to divide techniques applied to fault classification in power systems into the following three groups:

1. **Prominent techniques**: They used one of the following three approaches for feature extraction: wavelet transform, ANNs, and fuzzy logic.
2. **Hybrid techniques**: They are based on a combination of two or more approaches of the previous group (e.g., neuro-fuzzy technique, wavelet and ANN technique, wavelet and fuzzy-logic technique, wavelet and neuro-fuzzy technique).
3. **Modern techniques**: They are based on other AI techniques (e.g., SVM, GA, PCA, or DT) or use modern technologies to either measure signals or process extracted signal features (FPGA-based implementation, PMU-based protection scheme, pilot scheme).

Although this classification was proposed for techniques applied to transmission levels, it can also be considered for techniques to be applied to distribution levels.

Reference [9] divides the techniques into methods based on logic flow and methods based on AI techniques. The techniques of the first group use a tree-like logic flow with multiple criteria and values derived from the application of some feature extraction. The techniques of the second group use fuzzy inference systems, ANNs, SVMs, DTs, and other ML techniques.

**Fault location:** Fault location is a process aimed at locating the fault with the highest possibly accuracy [1]. When locating faults on a line consists of more than one section (i.e., a three-terminal or multi-terminal line), the faulty section has to be identified and a fault in this section has to be located. Even the location of temporary faults is important since the fault location can help to pinpoint the weak spots on the line so maintenance plans can be fixed for avoiding future problems. In a few words, a fault location method estimates the physical location of a fault by processing the voltage and current waveforms or the extracted features from those waveforms.

A fault-location function can be implemented into microprocessor-based protective relays, digital fault recorders, stand-alone fault locators, and post-fault analysis programs [1, 3]. A fault locator is a supplementary device with a fault-location algorithm for estimating the distance to the fault. Although fault locators and protective relays are closely related, there are some important differences between them: accuracy of fault location; speed of determining the fault position; speed of transmitting data from remote site; used data window; digital filtering of input signals and complexity of calculations [1]. Fault locators are used for accurately estimating the fault position and not only for indication of the general area where a fault occurred. Since protective relays perform their tasks online, high speed operation appears as a crucial requirement (i.e., the decision for tripping transmission lines has to be made in less than in one cycle of fundamental frequency). In contrast, the calculations of fault locators are performed in an off-line mode since the results of these calculations are for later use. This implies that the fault-location calculation can last seconds or even minutes. That is, the required high speed of protective relays imposes not too complex and too time consuming calculations; in contrast, fault-location calculations do not have such limitations. Consequently, the models of the power line and the fault in fault-location algorithms are usually more advanced than for relaying [1].

Distance relays are the protective relays more closely related to fault locators: they are designed for fast and reliable indication of the general area where a fault occurred. Generally, a pair of distance relays is used to protect a two-terminal line. Usually, they can communicate with each other, forming a pilot relay. The operation of a distance relay may be significantly influenced by the combined effect of load and fault resistance. The distance relay may not operate if the value of the fault resistance is too large. The value of the fault resistance may be particularly large for ground faults, which are the most frequent faults on overhead lines.

Methods based on visual inspection do not satisfy the requirements imposed on modern grids on fault location (i.e., automatic and fast location of the fault). The most popular fault location methods can be classified into the following categories [1]:

1. **Techniques based on fundamental-frequency currents and voltages**: These techniques assume that the calculated impedance of the faulted-line segment is a measure of the distance to fault. When applied to a two-terminal line, they can be classified considering the available measurements: waveforms from one or both ends; complete or incomplete measurements (voltage or current) from a particular line end. Methods using one-end impedance techniques do not need communication means and their implementation into digital protective relays or digital fault recorders is rather simple. However, the algorithms will be more accurate if information from the two line terminals is available; therefore, if communication channels are at the disposal, then two-terminal fault-location methods should be used since low-speed communications are sufficient. Besides, two-end techniques exhibit more accuracy without any assumption about external networks (i.e., impedances of the equivalent sources).
2. **Techniques based on traveling-wave phenomenon**: These techniques use voltage and current waves, traveling at the speed of light from the fault towards the line terminals; they can be very accurate, but also complex and expensive due to the required high sampling frequency.
3. **Techniques based on high-frequency components of currents and voltages**: These techniques are also complex and expensive since they require specially tuned filters for measuring high-frequency components.
4. **AI-based techniques**: Although ANN-based methods for fault location have been developed for more than forty years, it has been during the last two decades when a significant effort has been dedicated to fault-location techniques both in transmission and distribution networks using AI methods. A high number of review papers focused on this subject has been published during this period; see references [8, 9, 14, 15, 16, 22, 25, 29, 33, 36, 37, 41, 239, 240].

Two interesting guides on fault location methods in AC power systems are available in the technical brochure written by the CIGRE WG B5.52 [241] and in the IEEE standard C37.114-2014 [242].

Table 3 shows a selected list of review papers on fault diagnosis published since 2021.

The reminder of this section is dedicated to reviewing AI-based fault diagnosis techniques applied to power systems. The next two subsections are respectively dedicated to reviewing the work made on fault diagnosis of both transmission and distribution systems. The last subsection is dedicated to review works focused on DC systems.

*4.2. Fault Diagnosis Methods for Transmission Systems*

References [243-342] provide a list of papers that deal with AI-based techniques applied to detection, classification and location of faults in transmission systems, ordered by publication date. Note that, as for other power system studies and applications, AI-based techniques has been an option since 1980s. Due to room limitations, only a selection of those published since 2021 is presented in Table 4. Although the table does not provide too much information about the selected works, there are several conclusions that can be drawn:

1. There is a high number of AI techniques that can be useful to faults diagnosis of power systems. Remember that the topics covered in this survey are focused on faults/failures that can affect overhead lines and insulated cables only.
2. None of the selected papers deals with automatic AI-based fault diagnosis of transmission-level insulated cables.
3. Only a small percentage of papers deal with a complete fault diagnosis procedure (i.e., detection, classification and location). Actually, the fault location seems to be the task to which a lower number of papers has been dedicated.
4. Supervised ML techniques (i.e., NN, SVM, DT) are the most popular group of AI applications.

A high percentage of the selected works is based on either a two-terminal transmission line model or a rather small benchmark system, usually an IEEE test system. Only a few works used public datasets (e.g. datasets available in Kaggle); see, for instance, references [339] and [342].

**Table 3.** – Selected reviews on fault diagnosis in power systems using AI-based techniques.

| Ref. | Authors | Year | Topics |
|---|---|---|---|
| [42] | Daang et al. | 2024 | Fault detection in transmission lines |
| [41] | Kanwal and Jiriwibhakorn | 2024 | Fault detection, classification, and location in transmission lines |
| [40] | Shukla and Deepa | 2024 | Fault classification in transmission lines |
| [38] | Mishra, Gupta, and Yadav | 2023 | Travelling-wave based fault location methods in FACTS compensated transmission systems |
| [36] | Liu et al. | 2023 | Fault location in transmission lines |
| [35] | Shakiba et al. | 2023 | Fault detection, classification, and location in transmission lines |
| [34] | Jena et al. | 2023 | Fault detection, classification, and location in underground cables |
| [33] | Rezapour et al. | 2023 | Fault location in distribution grids |
| [8] | De La Cruz et al. | 2023 | Fault location in smart distribution grids and MGs |
| [48] | Baharozu et al. | 2023 | High impedance fault location |
| [32] | Shafiullah et al. | 2022 | Comparison of ML techniques for distribution grid fault analysis |
| [30] | Srivastava et al. | 2022 | Fault detection, isolation, and restoration in distribution grids |
| [29] | Stefanidou-Voziki et al. | 2022 | Fault classification and location in distribution grids |
| [26] | Panahi et al. | 2021 | Fault location in transmission grids |
| [25] | Dashti et al. | 2021 | Fault prediction and location in smart distribution grids and MGs |
| [24] | Vaish et al. | 2021 | Fault detection, isolation, and restoration in power systems |
| [23] | Nsaif et al. | 2021 | Fault detection in distribution grids with DG |
| [22] | Mukherjee et al. | 2021 | Fault detection, classification, and location in transmission lines |

DG: Distributed Generation, FACTS: Flexible AC Transmission System, MG: Microgrid, ML: Machine Learning.

As for software tools, most of the simulations required to obtain data were carried out using MATLAB/Simulink or an EMTP-like tool (e.g., ATP, PSCAD). This allowed authors to implement highly accurate models, although not all of them did choose such an option (e.g., a frequency-dependent model for representing overhead lines).

Additional works related to fault diagnosis of transmission-level systems in which new and emerging technologies (e.g., quantum computing, unmanned aerial vehicles-UAV) were adopted are mentioned in Section V.

*4.3. Fault Diagnosis Methods for Distribution Systems*

References [343-451] provide a list of papers dealing with AI-based techniques applied to detection, classification and location of faults in distribution systems. As for papers dealing with fault diagnosis of transmission systems, they are ordered by publication date.
Again, only a selection of those published since 2021 is presented in Table 5, from which some conclusions be drawn:
1. As for research related to transmission systems, a high number of AI techniques has also been applied to faults diagnosis of distribution systems.
2. Only a small percentage of works deals with a complete fault diagnosis procedure (i.e., detection, classification and location).
3. Supervised ML techniques are again the most popular option for authors interested in this field.

A high percentage of the selected papers is based on either a two-terminal line model or a rather small system. Although IEEE test systems were the option selected by many authors, many others used custom-made distribution system models. As for software tools, most of the simulations were carried out again using MATLAB/Simulink or an EMTP-like tool (e.g., ATP, PSCAD).

An interesting aspect has been the option selected by authors to perform the feature extraction task. Wavelet transform has been a popular approach but many works were based on different approaches, such as CAE, SIG or STFT (see Table 5).

**Table 4.** – Selected papers on AI-based fault diagnosis in transmission systems.

| Ref. | Authors | Year | Task | Technique |
|---|---|---|---|---|
| [342] | Anwar et al. | 2025 | Fault detection | RF, LSTM, and kNN |
| [341] | Nayak et al. | 2024 | Fault detection and classification | CWT and 2D-CNN |
| [340] | Wu et al. | 2024 | Fault classification and location | PNMCN |
| [339] | Turanli and Yakut | 2024 | Fault classification | 1D-CNN |
| [338] | Jia et al | 2024 | Fault location | CEEMDAN, MSA, and ConvGRU |
| [337] | de Alencar et al. | 2024 | Fault classification | ICA and CNN |
| [336] | Alhanaf et al. | 2024 | Fault detection and classification | Hybrid CNN-LSTM |
| [335] | Najafzadeh et al. | 2024 | Fault detection, classification, and location | WHO-RF/DT, ANFIS |
| [334] | Ukwuoma et al. | 2024 | Fault detection, classification, and location | MSAN, DGNN, and MLP |
| [333] | Chen and Liu | 2024 | Fault detection | FuzREANN |
| [332] | Mampilly and Sheeba | 2023 | Fault detection and classification | WT and ICNN-BOA |
| [331] | Bhattacharya and Nigam | 2023 | Fault detection and classification | RF, DT, XGB, LGBM |
| [330] | Altaie et al. | 2023 | Fault detection | Several ML techniques |
| [329] | Alhanaf et al. | 2023 | Fault detection, classification, and location | ANN, DNN |
| [328] | Khan et al. | 2023 | Fault classification and location | VAE and SVM, kNN, RF, DT |
| [327] | Biswas et al. | 2023 | Fault detection and classification | VMD and CNN |
| [326] | Zhang and Wang | 2023 | Fault classification | GLDA-CE |
| [324] | Goni et al. | 2023 | Fault detection and classification | ELM |
| [323] | Sahoo and Samal | 2023 | Fault detection and classification | DNN |
| [322] | Thomas et al. | 2023 | Fault detection and location | CNN |
| [321] | Rajesh et al. | 2022 | Fault detection and classification | TSVD-HUA-RPNN |
| [320] | Fahim et al. | 2022 | Fault detection and classification | DBN |
| [319] | Hong et al. | 2022 | Fault classification and location | CNN |
| [318] | França et al. | 2022 | Fault classification | MLPN, RBF, SVM, DT |
| [316] | Gutierrez-Rojas et al. | 2022 | Fault classification | DM-DFT and QARMA |
| [313] | Arranz et al. | 2021 | Fault location | ST and ANN |
| [312] | Fahim et al. | 2021 | Fault detection and classification | WT and CNSF |
| [311] | Rafique et al. | 2021 | Fault detection and classification | e2e learning and LSTM |
| [310] | Mukherjee et al | 2021 | Fault detection and location | PCA |
| [309] | Hassani et al. | 2021 | Fault classification | kGAN and kNN, SVM |
| [308] | Mukherjee et al. | 2021 | Fault classification | PCA |
| [307] | Belagoune et al. | 2021 | Fault detection, classification, and location | DRNN-LSTM |
| [306] | Srikanth and Koley | 2021 | Fault classification | ST and 3D CNN |
| [305] | Haq et al. | 2021 | Fault detection and classification | DWT and ELM |
| [304] | Vyasa et al. | 2021 | Fault detection and classification | DWT and ChNN |
| [303] | Han et al. | 2021 | Fault classification | GSV-CDA-CNN |

**Note**: The acronyms used in this table are defined after the Conclusion of this paper.

**Table 5.** – Selected papers on AI-based fault diagnosis in distribution systems.

| Ref. | Authors | Year | Task | Technique |
|---|---|---|---|---|
| [450] | Shafei et al. | 2024 | Fault detection, classification and location | PT-PFPT and CNN |
| [449] | Arsoniadis and Nikolaidis | 2024 | Fault location | WSN and SVM/ERM |
| [448] | Barkhi et al. | 2024 | Fault detection and classification in MGs | SVM |
| [447] | Krishnamurthy et al. | 2024 | Fault classification | RF |
| [446] | Yildiz and Abur | 2024 | Fault detection and location | CNN |
| [445] | Fan et al. | 2024 | Fault location | VGAE-GraphSAGE |
| [444] | Basher et al. | 2024 | Fault classification and location in MGs | DWT with DTE-LDA |
| [443] | Bhagwat et al. | 2024 | Fault detection, classification and location | Customised ANN |
| [442] | Awasthi et al. | 2024 | Fault classification | kNN |
| [441] | Liang et al. | 2024 | Fault location | Multi-head GAT |
| [440] | Zhou et al. | 2024 | Fault classification and location | CNN |
| [439] | Cieslak et al. | 2024 | HIF classification in MGs | TNN |
| [438] | Li | 2024 | HIF detection and location | CNN-LSTM |
| [436] | Mampilly and Sheeba | 2024 | Fault detection and classification in MGs | EWT-HCRNN-POA |
| [435] | Awasthi et al. | 2024 | Fault classification | Shallow ANN |
| [434] | Bhatnagar et al. | 2024 | Fault detection and classification | CNN-LSTM-AM |
| [433] | Mbey et al. | 2023 | Fault detection and classification | LSTM-ANFIS |
| [432] | Mirshekali et al. | 2023 | Fault location | Spectrogram and CNN-CN |
| [425] | Yang and Yang | 2023 | Fault classification | STFT and CNN-FDTW |
| [424] | Kurup et al. | 2023 | Fault detection and classification | CNN and SVM |
| [423] | Mo et al. | 2023 | Fault classification and location | Super-resolution and GNN |
| [422] | Haydaroğlu and Gümüş | 2023 | Fault detection | Cauchy-M and RVFLN |
| [421] | Rizeakos et al. | 2023 | Fault classification and location | CWT and CNN |
| [420] | Yuan and Jiao | 2023 | Fault detection | Hybrid CNN-LSTM |
| [415] | Dashtdar et al. | 2023 | Fault location | GA |
| [414] | Hu et al. | 2023 | Fault classification and location | STGCN |
| [413] | da Silva Santos et al. | 2022 | Fault detection and classification | DWT and FIS |
| [412] | Moloi et al. | 2022 | Fault classification and location | WPD and SVM |
| [411] | Ahmadipour et al. | 2022 | Fault detection and classification in MGs | MODWPT-ALPSO-SVM |
| [410] | Granado Fornás et al. | 2022 | Fault detection and classification | TDR, GAF, and GAN |
| [406] | Rai et al. | 2022 | HIF detection and classification | TNN-CNN |
| [401] | Carvalho et al. | 2022 | HIF classification | HOS-FDR with ANN |
| [399] | Mirshekali et al. | 2022 | Fault location | NCFS and SVM |
| [398] | Swaminathan et al. | 2021 | Fault classification and location in cables | CNN-LSTM |
| [397] | de Freitas and Coelho | 2021 | Fault location | GNN |
| [396] | Gilanifar et al. | 2021 | Fault classification | MTLS-LR |
| [394] | Yu et al. | 2021 | Fault location | SIG-CNN |
| [392] | Okumus and Nuroglu | 2021 | Fault location | WT-FWHT and RF |
| [390] | Baloch and Muhammad | 2021 | Fault detection and classification in MGs | HT, LR and AB |

**Note**: The acronyms used in this table are defined after the Conclusion of this paper.

*4.4. Fault Diagnosis Methods for DC Systems*

Although there are significant differences between the technologies currently implemented to protect DC lines and cables with respect to those implemented for protecting AC systems, fault diagnosis techniques applied to both types of systems exhibit some similarities. The protection of DC systems has been the subject of several recent review papers [452-455]. Likewise, fault detection, identification and location techniques applied to DC systems have also been reviewed in some recent works [455-458]. Reference [455] provides an extensive review of techniques implemented for protection and fault diagnosis of HVDC systems, including the technologies used in their design. As with AC systems, a significant effort has been made during the last decades for analyzing the potential that AI-based techniques offer in fault diagnosis of DC systems; see [459-490]. Table 6 provides a selection of papers published since 2021. Note that the list of references mentioned in the table includes papers related to any voltage level and covers the most common technologies (i.e., CSC, VSC, MMC). The test system configurations (an aspect not mentioned in the table) cover from two-terminals [459, 460, 475, 479, 488, 490] to multi-terminal DC systems [460, 463, 474, 478, 485, 486], as well as low-voltage DC microgrids [467, 472, 473] and systems based on the CIGRE benchmark model [459]. As for the techniques, there are examples from all AI subfields, but it seems that those based on neural networks are the most popular.

**Table 6.** – Selected papers on AI-based fault diagnosis in DC systems.

| Ref. | Authors | Year | Task | Techniques |
|---|---|---|---|---|
| [490] | Liu et al. | InPress | Traveling wave fault location in MMC-based HVDC systems | GWO-VMD |
| [489] | Fayazi et al. | 2025 | Fault detection and classification in parallel HVAC/HVDC transmission lines | DT and FEI |
| [488] | Akbari and Shadlu | 2024 | Fault detection, classification and location in VSC-based HVDC systems | PCA-DWT and ANFIS |
| [487] | Pragati et al. | 2024 | Fault detection and classification in VSC-based HVDC systems | ST-RNN and TEO |
| [486] | Yousaf et al. | 2024 | Fault detection in MMC-based HVDC systems | DWT and Enhanced ANN with Bagging |
| [485] | Yousaf et al. | 2024 | Fault detection in MMC-based HVDC systems | LSTM-DWT |
| [484] | Yu et al. | 2024 | Fault location in DC distribution systems | Multivariate information fusion |
| [483] | Deb and Jain | 2024 | Fault detection and classification in low-voltage DC microgrids | BEL and cosine kNN |
| [482] | Salehimehr et al | 2024 | Fault detection and location in low-voltage DC microgrids | CS-RT and LSTM |
| [480] | Hameed et al. | 2024 | Fault detection and classification in MMC-based HVDC systems | HHO and ANN |
| [479] | Jawad and Abid | 2023 | Fault detection in VSC-based HVDC systems | ACO-DWT and ANN |
| [478] | Gnanamalar et al. | 2023 | Fault detection, classification and location in VSC-HVDC systems | HHT plus CNN-SVM |
| [477] | Psaras et al. | 2023 | Fault location in HVDC systems | GA in frequency domain |
| [476] | Jawad and Abid | 2022 | Fault detection in VSC-based HVDC systems | GWO and ANN |
| [475] | Ghashghaei and Akhbari | 2021 | Fault detection and classification in CSC-HVDC systems | SVM and KNN |
| [474] | Yang et al. | 2021 | Fault detection and location in MMC-based HVDC systems | CWT and Deep-RNN |
| [473] | Roy | 2021 | Fault detection and location in HVDC systems | DOST-PNN and FDST-BPNN |
| [472] | Wang et al. | 2021 | Fault location in VSC-based HVDC transmission lines | VMD-TEO and CNN-LSTM |
| [471] | Ye et al. | 2021 | Fault location in MMC-based HVDC systems | WT and DBN |
| [470] | Wu et al. | 2021 | Fault location in MMC-based HVDC systems | SVM-CEEMDAN |

**Note**: The acronyms used in this table are defined after the Conclusion of this paper.

## V. Discussion

There are many aspects related to the main topics covered by this survey that are worth mentioning or clarifying. Some of them are discussed below.

1. This paper holds the concept Artificial Intelligence in its title. Interestingly, future developments in this field, namely in generative artificial intelligence (GAI) [491], might make unnecessary the effort to create such type of documents since it cannot be discarded that a survey paper like the current one could be easily generated by taking advantage of future developments of AI.
2. The term GAI is applied to models that are used to generate new data: the models learn from training data and generate new data similar to what they have been trained on. Large Language Models are perhaps the most popular generative AI models; they are trained using a self-supervised learning approach and a vast amount of text [492, 493]. A review of existing studies and potential applications of GAI models in the energy sector was presented in [494]. The potential in power grid operations has been analyzed in [495]. Among others, GAI models have already been applied to power grid visualization of large scale transmission networks [496] or to generate realistic augmented datasets [497].
3. What is AI and what is not is another important aspect. Although part of Section 2 was dedicated to clarifying this aspect, the fact is that AI concepts has been classified using different approaches in many previous works. For instance, some authors include game theory as a subfield of AI; see, for instance, [207]. Neither game theory nor other popular approaches (i.e. multi-agent system) were considered in this survey.
4. The availability of massive datasets and open-source code, the development of efficient algorithms, and the continuous improvement of computing power are some of the aspects that have made possible the application of AI algorithms.
5. The number of papers related to fault diagnosis (i.e., detection, classification and location of faults) in power systems is about several thousand. Although this review has only covered papers related to AI-based techniques, some selection has been unavoidable. Obviously, it is debatable the way in which the papers included in this survey have been selected.
6. Unmanned aerial vehicles (UAVs), also known as drones, have emerged as an option for inspection of overhead lines and for detection of faults [498, 499]. UAV-mounted sensors can be a data source for accurate visual and thermal inspection of overhead lines. There is an increasing interest in the application of AI-based techniques to the detection and classification of overhead line faults using data from UAVs [500-506].
7. AI-based methods exhibit some potential for improving accuracy and adaptability to diverse fault conditions; however, their practical implementation is challenging. Consider, for example, a fault location scheme using a traveling wave-based approach combined with ANNs; such technique requires extensive training data and high computational burden, in addition to a continuous adaptation to varying system configurations. This can be especially difficult for distribution networks with high renewable penetration since they require innovative fault location techniques that can handle bidirectional power flows.
8. The practical implementation of AI-based techniques for detecting, classifying and locating faults will be parallel to advances in software and hardware. Training data for supervised ML techniques can be easily derived from computer simulations, which should be carried out using sophisticated software tools and very accurate power system component models. A similar conclusion can be derived from hardware implementation: AI-based techniques require powerful and flexible microprocessors that could be easily reprogrammed considering the experience obtained from both the actual power system and its computer representation. Real-time simulation platforms will be of much help in deploying the new techniques. Consider that many works have been based on results derived from rather small test systems, and many authors did not include the representation of instrument transformers (i.e., current and voltage transformers) in the system models. Therefore, it is advisable to be careful about some conclusions.
9. An aspect that has to be considered for selecting an adequate technique is the data that can be available, and this will depend on the various technologies installed in the system under study. The modern SG offers sufficient availability of data for the implementation of an accurate AI-based fault diagnosis technique. That is, an aspect that can affect a fault diagnosis scheme based on an AI technique is the monitoring system implemented in the actual power grid. The grid becomes smarter as the number of monitoring nodes increases. The way in which a practical fault location scheme will be developed and implemented depends on this aspect and on the skills of the available communication system. From a theoretical point of view estimating the faulted section in a distribution system would be a very easy task if a monitor has been installed in each grid node and a low-latency communication network is available; in practice, such an ideal scenario would be too expensive and hardly justifiable.
10. The faulty section of a power system can be easily and quickly estimated by means of an expert system if powerful monitoring and low-latency communication systems have been installed in the grid. A combination of an ES and a ML-based technique might be a practical solution for detecting, classifying and location faults in smart grids.

11. As discussed in subsection 4.1, a fault-location function can be part of a microprocessor-based protective relay, and a fault locator can be a supplementary device that can estimate the fault location in an overhead line or an insulated cable. Presently, a fault locator is also a device manually handled by maintenance crews to estimate the location of faults, mainly in underground cables.

12. Although most (if not all) of the works reported in this survey are based on computer simulation, the survey has not covered some important aspects such as the applied software simulation tools or the sources from which data for training ML techniques come from. As for simulation tools, it has already been mentioned that most works are based on MATLAB and EMTP-like tools. In general, data for training ML-based approaches come from simulations carried out by the authors; however, it is worth mentioning that some dataset repositories are available for helping researchers in the development of predictive models. Some works that could be useful to readers interested in this subject were presented in references [507-511].

13. Although a significant effort has been made to date and very useful experience is already available on the application of AI-based techniques to detect, classify and locate faults in lines and cables, it is not easy to select the best combination of techniques. A very good method for locating faults in a distribution system based on a rather limited system model should not be selected as a winner: although the combination of techniques selected for each task (i.e. feature extraction, detection, classification, location) can be useful for future work, it could also exhibit poor performance when using a more accurate and sophisticated system model.

14. The number of AI algorithms applied to power system studies is steadily increasing. The list of the latest techniques include transfer learning, graph learning, deep attention mechanism, deep reinforcement learning, or physics-guided neural networks. Some of these developments address some limitations of neural networks (e.g., overfitting, low data efficiency, low adaptivity, or physical inconsistency).

15. Quantum computing is an emerging technology that will be extremely useful in AI-based applications for which high performance computing power is a requirement [512-515]. Its application to power system analysis was reviewed in references [516-517]. Although not much experience is available to date, some interesting works on fault diagnosis of power systems has already been presented [518-521].

16. The YOLO algorithm is an interesting approach that has recently been applied to fault diagnosis of power systems. YOLO is an acronym that stands for *You Only Look Once*; it is an object detection algorithm based on CNN introduced by Redmon et al. [522]. Since its presentation, YOLO has undergone many improvements; it is currently one of the most popular object detection frameworks, and exhibits a remarkable performance of speed and accuracy [523]. During the last years, various versions of this algorithm have been applied to detection, classification and location of faults in both transmission and distribution levels; see, for instance [524-530].

17. Satellite applications can support power system analysis and operation in many areas (e.g., asset management, two-way communication between smart meters and grid operators, prediction of consumption/generation peaks, use of virtual power plants, integration of electric vehicles into the electricity grid, cybersecurity). In power system protection and fault location, real-time clock signal synchronization is a crucial aspect that is usually implemented by means of the GPS, a satellite-based navigation system [531]. Travelling wave based methods requiring two-terminal data usually employ the GPS to keep the measurements synchronized. The advantages of using a time signal synchronization was analyzed in [532], a paper that assumes GPS synchronization based on the IEEE Std 1588 protocol [533]. The need for synchronization using GPS is also discussed in [534], which presents a short story of PMUs, synchrophasor technology and wide area monitoring systems (WAMS), a technology to monitor power system dynamics in real time. Remember that PMUs are devices that provide harmonized measurements of real-time phasors of voltages and currents, being the harmonization achieved by sampling voltage and current waveforms using timing signals from the GPS. For the application of techniques in which the GPS time-synchronized signals are used, see [388, 396, 399, 438, 439, 535-538].

18. Since AI systems are prone to cyberattacks, risk evaluation can be crucial before their implementation. In complex transmission and distribution grids, associated with the vulnerability of SCADA systems and communication networks, which interconnect countless smart devices (meters, sensors, etc.), cybersecurity becomes a critical issue. Attacks to SG equipment (e.g., data breaches, data manipulation, unauthorized access, denial of service, man-in-the-middle, false data injection, malware introduction, etc.) can cause large-scale damage. Some recent works, see [539, 540], offer a broad perspective on these risks and alert to the need for investment in strengthening security and mitigating potential damage. Besides, with their increasing autonomy, AI systems are acquiring skills that can be dangerous even for themselves; therefore, risk evaluation of their capabilities before implementation is becoming another critical issue [541].

19. Federated Learning (FL) is another machine learning framework that uses multiple data sources to collaboratively train a model without sharing their raw data: instead of collecting all data at a unique node, each node computes model updates on its own local dataset and shares these updates with the central coordinating node. This decentralized approach allows keeping sensitive information private, so it is particularly interesting in sectors where data privacy and security are critical. An aspect that distinguishes FL from traditional distributed learning is the nature of the data, since FL often deals with heterogeneous data. For an overview of methods and applications of FL, see references [542-545]. The application and benefits of FL in power and energy systems has been reviewed in [546, 547]. Since modern smart grids are equipped with a myriad of sensors that monitor voltage, current, and frequency, FL models can be trained to detect anomalies (e.g, faults) in real time; therefore, FL is especially adapted to analyze grid reliability by using decentralized data as the grid becomes smarter and more interconnected. Some application of FL to fault diagnosis studies has been presented in [548, 549].

## VI. Conclusion

This paper has presented a bibliographical survey on AI-based techniques for detecting, classifying and locating faults in overhead lines and insulated cables operating at any voltage level, including works related to microgrids and DC systems. Although a practical and efficient implementation of a complete fault diagnosis does not seem to be possible in the short term, it is evident that an increasing effort will continue being dedicated to analyzing the potential that the new techniques offer in power system analysis and design. The number of works has significantly increased during the last few years, new techniques are being tested for every fault diagnosis task. The last progress reported on AI technologies and their applications supports this conclusion.

**List of Acronyms**

| | |
|---|---|
| AC: alternating current | HIF: high impedance fault |
| ACO: ant colony optimization | HLLE: Hessian locally linear embedding |
| AB: AdaBoost | HOS: higher-order statistics |
| AI: artificial intelligence | HT: Hilbert transform |
| ALPSO: augmented Lagrangian particle swarm optimization | HUA: human urbanization algorithm |
| AM: attention mechanism | HVAC: high voltage alternating current |
| ANFIS: adaptive neuro-fuzzy inference system | HVDC: high voltage direct current |
| ANN: artificial neural network | ICNN: improved convolution neural network |
| BEL: bagged ensemble learner | KMC: K-means clustering |
| BOA: bees optimization algorithm | kGAN: knockoff generative adversarial network |
| BPNN: backpropagation neural network | kNN: k-nearest neighbour |
| CAE: convolutional auto-encoder | LDA: linear discriminant analysis |
| CDA: cross-domain adaption | LGBM: light gradient boosting machine |
| CE: characteristic entropy | LR: logistic regression |
| CEEMDAN: complete ensemble empirical mode decomposition with adaptive noise | LSTM: long short-term memory |
| ChNN: Chebyshev neural network | MG: microgrid |
| CN: Capsule network | MIF: multivariate information fusion |
| CNN: convolutional neural network | ML: machine learning |
| CNSF: capsule network with sparse filtering | MLP: multi-linear perceptron network |
| ConvGRU: convolutional gate recurrent unit | MLPN: multi-layer perceptron neural network |
| CS: compressed sensing | MMC: modular multilevel converter |
| CSC: current source converter | MODWPT: maximal overlap discrete wavelet packet transform |
| CWT: continuous wavelet transform | MRA: multiresolution analysis |
| DBN: deep belief network | MSA: mantis search algorithm |
| DC: direct current | MSAN: multi-scale attention network |
| DER: distributed energy resource | MTLS-LR: multi-task latent structure learning |
| DFT: discrete Fourier transform | NCFS: neighborhood component feature selection |
| DG: distributed generation | NN: neural network |
| DGNN: deep graph neural network | PCA: principal component analysis |
| DL: deep learning | PFPT: Piecewise Function Put Together algorithm |
| DM-DFT: delta method discrete Fourier transform | PMU: phasor measurement unit |
| DNN: deep neural network | PNMCN: pose normalized multioutput convolutional nets |
| DOST: discrete orthonormal S-transform | PNN: probabilistic neural network |
| DRNN: deep recurrent neural network | POA: pelican optimization algorithm |
| DT: decision tree | PSO: particle swarm optimization |
| DTE: decision tree ensemble | PT: Park's transformation |
| DWT: discrete wavelet transform | QARMA: quantitative association rule mining algorithm |
| e2e: end to end learning | QL: Q-learning |
| ELM: extreme learning machine | RBF: radial basis function neural network |
| EMS: energy management services | RF: random forest |
| ERM: ensemble regression mode | RMS: root mean square |
| ES: expert system | RNN: recurrent neural network |
| EWT: empirical wavelet transform | RPNN: recurrent perceptron neural network |
| FDIR: fault detection, isolation, and service restoration | RT: regression tree |
| FDR: Fisher's discriminant ratio | RVFLN: random vector functional link network |
| FDST: Fast discrete S-transform | SA: simulated annealing |
| FDTW: fast dynamic time warping | SCADA: supervisory control and data acquisition |
| FEI: fault energy index | SG: smart grid |
| FFT: fast Fourier transform | SIG: signal to image |
| FIS: fuzzy inference system | ST: S-transform/Stockwell transform |
| FL: federated learning | STFT: short-time Fourier transform |
| FLISR: fault location, isolation, and service restoration | STGCN: spatiotemporal graph convolutional network |
| FLR: fuzzy linear regression | SVM: support vector machine |
| FPGA: field programmable gate array | TDR: time-domain reflectometry |
| FPI: fault passage indicator | TEO: Teager energy operator |
| FT: Fourier transform | TNN: transformer neural network |
| FuzREANN: fuzzy reinforcement encoder adversarial neural networks | TS: tabu search |
| FWHT: fast Walsh Hadamard transform | t-SNE: t-distributed stochastic neighbor embedding |
| GA: genetic algorithm | TSVD: truncated singular value decomposition |
| GAF: Gramian angular field transform | UAV: unmanned aerial vehicle |
| GAI: generative artificial intelligence | VAE: variational encoder |
| GAN: generative adversarial network | VGAE: variational graph auto-encoder |
| GAT: graph attention network | VSC: voltage source converter |
| GLDA: global and local discriminant analysis | VMD: variational mode decomposition |
| GMM: Gaussian mixture model | WAMS: wide area monitoring system |
| GNN: graph neural network | WPD: wavelet packet decomposition |
| GPS: global positioning system | WSN: wavelet scattering network |
| GSV: gradient similarity visualization | WT: wavelet transform |
| GWO: grey wolf optimization | XAI: explainable artificial intelligence |
| HCRNN: hybrid convolutional recurrent neural network | XGB: XGBoost |
| HHO: Harris Hawks optimization | YOLO: you only look one |
| HHT: Hilbert–Huang transform | |